\documentclass[a4paper]{article}

\usepackage{makecell}
\usepackage{xcolor}
\usepackage{url}
\usepackage{booktabs}
\usepackage{multirow}
\usepackage{INTERSPEECH2019}
\usepackage{amssymb}% http://ctan.org/pkg/amssymb
\usepackage{pifont}% http://ctan.org/pkg/pifont
\newcommand{\cmark}{\ding{51}}%
\newcommand{\xmark}{\ding{55}}%

\newcommand{\newpara}[1]{\vspace{4pt}\noindent\textbf{#1}}

\usepackage[implicit=false]{hyperref}

\title{Spot the conversation: speaker diarisation in the wild}
\name{Joon Son Chung$^{1,2}$*, Jaesung Huh$^{1,2}$*, Arsha Nagrani$^{1}$*, Triantafyllos Afouras$^{1}$, Andrew Zisserman$^{1}$\thanks{\hspace{-12pt}* These authors contributed equally to this work.}}
\address{$^{1}$Visual Geometry Group, Department of Engineering Science, University of Oxford, UK \\$^{2}$Naver Corporation, South Korea}
\email{
\url{http://www.robots.ox.ac.uk/~vgg/data/voxconverse}}%\\
\begin{document}

\maketitle
\begin{abstract}
The goal of this paper is speaker diarisation of videos collected `in the wild'. 

We make three key contributions. 
First, we propose an automatic audio-visual diarisation method for YouTube videos. 
Our method consists of active speaker detection using audio-visual methods and speaker verification using self-enrolled speaker models. Second, we integrate our method into a semi-automatic dataset creation pipeline which significantly reduces the number of hours required to annotate videos with diarisation labels. Finally, we use this pipeline to create a large-scale diarisation dataset called \texttt{VoxConverse},  collected from `in the wild' videos, which we will release publicly to the research community. Our dataset consists of overlapping speech, a large and diverse speaker pool, and challenging background conditions.  
% First, we propose an automatic and scalable audio-visual diariation pipeline based on computer vision techniques to create a large diarisation dataset from YouTube videos. 
% Our pipeline is based on active speaker detection using audio-visual methods and speaker verification using self-enrolled speaker models. Second, we introduce an iterative strategy of manually verifying the diarisation labels predicted from the automatic pipeline, then optimising the automatic pipeline using this manually corrected data. The increase in prediction accuracy from the iterative process significantly accelerates the verification. Finally, we release this large-scale diarisation dataset called \texttt{VoxConverse},  collected from `in the wild' videos, publicly to the research community. Our dataset consists of overlapping speech, a large and diverse speaker pool, and challenging background conditions.  
\end{abstract}
\vspace{10pt}

%%% ========== ========== ==========
%%% Intro
%%% ========== ========== ==========

\section{Introduction}
Speaker diarisation is the challenging task of breaking up multi-speaker video into homogeneous single speaker segments, effectively solving \textit{“who spoke when”}. Beyond being an interesting research problem in itself, it is also a valuable pre-processing step for a number of applications, including speech-to-text. 

While state-of-the-art diarisation systems perform remarkably well for speech from constrained domains (e.g.\ conversational
telephone speech~\cite{sell2014speaker,zhu2016online,garcia2017speaker,zhang2019fully} or meeting speech~\cite{yella2013improved}), this
success does not transfer to more challenging conditions found in online videos `in the wild'. The challenges here include the lack of a fixed domain (videos can be from talk shows, news broadcasts, celebrity interviews, home vlogs), a large number of speakers (some of whom are off-screen), short rapid exchanges with cross-talk, and background degradation consisting of channel noise, laughter and applause. 

These conditions make manual annotation of online videos a daunting task for human annotators, leading to a dearth of large-scale public diarisation datasets of unconstrained speech. While large-scale evaluations are held regularly by
the National Institute of Standards in Technology (NIST-RTE), these are limited to constrained audio-only datasets, which are not freely available to the research community (Table \ref{table:existingdata}).

To attempt to remedy some these issues, the DIHARD challenges~\cite{sell2018diarization, ryant2019second} were introduced in 2018. These are valuable annual challenges that cover 11 different data domains, including mother-child conversations, meetings and courtroom settings. One of these domains is also web videos, however there is a limited amount of data (only 2 hours). The datasets are also audio-only, and are only available to challenge participants (not released freely to the research community).

A large-scale diarisation dataset of videos `in the wild' would encourage the development of new audio-visual diarisation techniques that deal with unconstrained conditions. Inspired by the recent success of automatic audio-visual dataset creation pipelines (VoxCeleb~\cite{Nagrani17,Chung18a,nagrani2020voxceleb}, VGGSound~\cite{vedaldi2020vggsound}), we propose a scalable, audio-visual method for speaker diarisation in web videos. Our method relies heavily on the recent successes of active speaker detection~\cite{Chung16a} and face and speaker verification~\cite{Parkhi15,Schroff15,Xie19a}. We then integrate this method into a semi-automatic dataset creation pipeline -- consisting of both automatic annotation and manual verification. 
We use this pipeline to curate \texttt{VoxConverse}, a challenging and diverse speaker diarisation dataset from `in the wild' videos. 

\begin{figure*}[!htb]
 \centering 
 \includegraphics[width=1\linewidth]{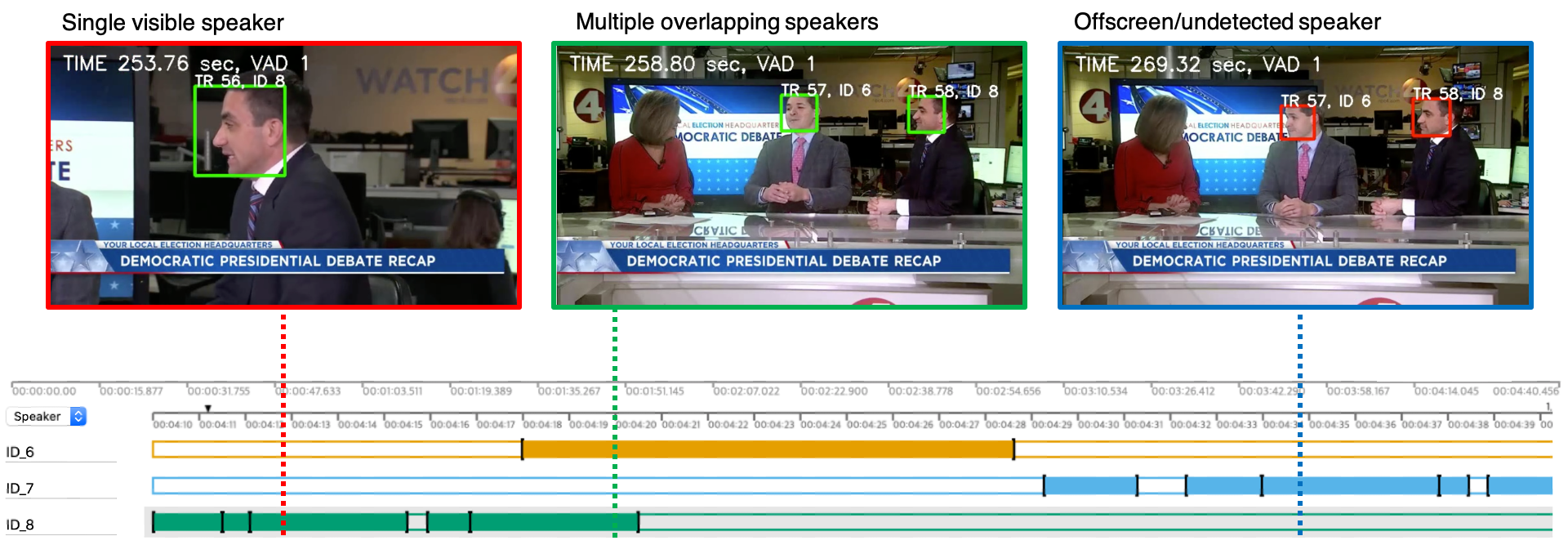}
 \caption{Output of our automatic audio-visual diarisation method. Green squares on the images represent face detections with positive ASD output, red squares represent face detections with negative ASD output. The identities are labelled as ID\_6, ID\_7 and ID\_8, and speaker timelines show when each identity is speaking. For clarity, we only show 3 frames from the video. Our method elegantly deals with visual speakers, overlapping speech and undetected/off-screen speakers. 
 % ** will be replaced with higher resolution figure**
 }
 \label{fig:anntool}
 \end{figure*}

% Using a semi-automatic process consisting of automatic annotation stage and manual verification stage, we generate a challenging and diverse speaker diarisation dataset from `in the wild' videos. 
Our automatic diarisation method exploits the following three key ideas; Firstly, the speech for on-screen identities can be accurately segmented automatically using active speaker detection and then identified using face recognition, the core basis for the VoxCeleb pipeline~\cite{nagrani2020voxceleb}. Second, there has been great progress in creating audio-visual speech enhancement models~\cite{Afouras18,Ephrat18,afouras2019my}, which separate overlapping speech into single speaker streams. Given the amount of cross-talk and background noise in web videos, we use this model to better isolate and identify speaker identities. The above two ideas allow us to accurately identify and isolate speech for on-screen speakers. Finally, to accurately recognise \textit{off-screen} speakers, we utilise state of the art speaker recognition embeddings that verify identities from audio alone (Figure \ref{fig:anntool}). 

Concretely, we make the following three contributions: (i) We create an automatic audio-visual diarisation method using active speaker detection, face recognition, speech enhancement and audio-only speaker recognition.
(ii) We integrate our method into a semi-automatic dataset creation pipeline which consists of human annotation and automatic diarisation. Our pipeline is scalable, and significantly reduces the number of hours required to annotate videos.
(iii) We use this pipeline to curate \texttt{VoxConverse}, a challenging `in the wild' audio-visual diarisation dataset. We compare our audio-visual diarisation method to existing audio-only baselines on our dataset, and show that large performance gains can be obtained from integrating visual information.
% \newpara{Notes to reviewers.}
% The dataset currently contains 20 hours of annotated video, but we are the process of scaling up. The pipeline is fully scalable and is effective across a range of domains. The final data statistics will be included in the camera ready version of the paper (if accepted), and will be publicly released after the second VoxCeleb Speaker Recognition Challenge in October 2020.

%%% ========== ========== ==========
%%% Related works
%%% ========== ========== ==========

\section{Related works}
 
Speaker diarisation has been an active field of research for many years, but remains one of the most challenging tasks in speech processing. Deep learning techniques have not been applied to speaker diarisation to the same degree that they have for other tasks, partially due to the lack of end-to-end models for diarisation, but also due to the lack of diverse, large-scale datasets like ImageNet~\cite{Deng09} and VoxCeleb~\cite{Nagrani17}. 

Much of the progress in the field has been driven by a series of NIST Rich Transcription challenges (NIST-RTE), which focuses almost solely on the meeting domain. The series also proposed the diarisation error rate (DER) as an evaluation metric for speaker diarisation, which is now used as the primary metric across all domains and evaluations.
Research into speaker diarisation has largely evolved independently for different domains, with broadcast news~\cite{tranter2004speaker}, telephone speech~\cite{canavan1997callhome}, and meetings~\cite{janin2003icsi,carletta2005ami} being the most popular domains. For each domain, specific datasets have been introduced and used, all created by manual annotation. We provide a summary in Table~\ref{table:existingdata}.

The DIHARD series of challenges \cite{sell2018diarization, ryant2019second} were introduced to overcome the domain dependency in the field -- the data consists of recordings from different conversational domains, including audiobooks, broadcast interviews, child speech and so on. The evaluation conditions are challenging, and even the best performing systems score relatively high diarisation error rates of around 20\% with ground truth voice activity detector (VAD), and 30\% with system VAD. The annotation is performed with very fine granularity, which allows evaluation without a forgiveness collar.
Barring DIHARD, all other datasets and evaluations include a generous forgiveness collar and exclude overlapping speech from scoring. Inspired by DIHARD, we annotate overlapping speech in \texttt{VoxConverse} and include it in evaluation. For almost all existing datasets, annotation is done manually and solely using the audio. Annotation without visual information is challenging, particularly when the number of speakers is large, since it is easy to be confused between voices without the additional identity redundancy provided by the face. Unlike other works, our dataset creation pipeline is semi-automatic, scalable and audio-visual.

\begin{table}[h!]
\centering
\renewcommand\arraystretch{1.1}
\setlength{\tabcolsep}{4pt}
\caption{
Comparison to existing speaker diarisation datasets. 
{\bf Acou. Cond.:} Acoustic conditions;
% {\bf Spks:} Speakers; 
\textbf{Ann. Method:} Annotation Method;
$\dag$: Fisher English Training Speech part I and II.
}
\label{table:existingdata}
\footnotesize
\begin{tabular}{ l  r  r  r  }
\toprule
~~~~~~~~  \textbf{Name} 							         & \textbf{Acou. Cond.} & \textbf{Free}  & \textbf{Ann.~Method} \\ 
\midrule
% 2000 NIST SRE  &  & \xmark     &    Manual  \\ 
2005 NIST RTE  & Meetings & \xmark     &   Manual  \\ 
CALLHOME~\cite{canavan1997callhome} & Telephony  & \xmark  &   Manual   \\ 
 AMI Meeting Corpus~\cite{carletta2005ami}  & Meetings  & \cmark &    Manual \\ 
 ICSI Meeting Corpus~\cite{janin2003icsi}        & Meetings   & \cmark     &    Manual\\ 
 Fisher$\dag$ I and II~\cite{cieri2004fisher} & Telephony &  \xmark &    Manual    \\ 
% Fisher ET$\ddagger$ II~\cite{cieri2004fisher}  & Telephony & \xmark & Manual       \\ 
DIHARD \cite{sell2018diarization, ryant2019second}   &  Mixed &  \xmark &  Manual		 \\
\midrule 

 \textbf{VoxConverse}							       & Multi-media   & \cmark &  Semi-automatic\\ 
 \bottomrule
\end{tabular} 
\normalsize
\end{table}

%%% ========== ========== ==========
%%% Dataset
%%% ========== ========== ==========

\section{Dataset description}
The development set of \texttt{VoxConverse} consists of 216 multispeaker videos covering 1,218 minutes with 8,268 speaker turns annotated. 
The test set contains approximately 232 videos covering 2,612 minutes.
The statistics of the dataset can be seen in Table \ref{table:data}. 

Videos included in the dataset are shot in a large number of challenging multi-speaker acoustic environments, including political debates, panel discussions, celebrity interviews, comedy news segments and talk shows. This provides a number of background degradations, including dynamic environmental noise with some speech-like characteristics, such as laughter and applause. Our dataset is audio-visual, and contains face detections and tracks as part of the annotation. 

The videos in the datasets consist of quick, short speech segments. On average, 91\% of the video time contains speech, and 3--4\% of this contains speech where one speaker overlaps with another speaker. The overlap percentage varies between videos; one video for example has an overlap percentage of 29.8\%. Videos vary in length from 22 seconds to 20 minutes. Unlike other domains such as telephony, each video has on average between 4 and 6 speakers, with one video in the dataset having 21 speakers. 

% The dataset will be made freely available to the research community. 

\begin{table*}[h!]
\centering
\renewcommand\arraystretch{1.2}
\caption{\texttt{VoxConverse}
dataset statistics.
% for the dev set. 
% The statistics of the test set will be released after VoxSRC 2020.
Entries that have 3 values are reported as min/mean/max. \textbf{\#~spks:} Number of unique speakers per video. \textbf{\#~mins:} Total number of minutes in the dataset. \textbf{video durations (s):} Length of videos in seconds. \textbf{speech \%:} Percentage of video time that is speech. \textbf{overlap \%:} Percentage of speech per video when 2 or more speakers overlap.}
\label{table:data}
\footnotesize
\begin{tabular}{ l  r  r  r  r r r r r}
\toprule
 \textbf{set} & \textbf{\# videos} & \textbf{\# mins}  & \textbf{\# spks}  & \textbf{video durations (s)} & \textbf{speech \%} & \textbf{overlap \%} \\  % & \textbf{\# speaker turns}
\midrule 
%  Dev  & 105   & 601 & 1 / 4.5 / 12 & 4,241   & 22.0 / 347.2 / 1014.8       &    68.0 / 94.1 / 99.6    & 0 / 3.4 / 27.4\\ 
%  Dev   & 113   & 604 & 1 / 4.6 / 19 & 3,688 & 26.3 / 328.0 / 1097.4     &    11.1 / 93.4 / 99.8    & 0 / 2.3 / 16.6 \\ 
 Dev   & 216   & 1,218 & 1 / 4.5 / 20  & 22.0 / 338.2 / 1097.4     &    10.7 / 93.2 / 99.8    & 0 / 3.8 / 28.7 \\  % & 8,268
 Test   & 232   & 2,612 & 1 / 6.5 / 21 & 26.0 / 675.6 / 1200.0     &    46.9 / 89.6 / 100    & 0 / 3.1 / 29.8 \\  %  & x,xxx
%  Test   & \multicolumn{7}{c}{\textit{To be released in October 2020}} \\ 
 \bottomrule
\end{tabular} 
\normalsize
\end{table*}

%%% ========== ========== ==========
%%% Pipeline
%%% ========== ========== ==========

\section{Dataset collection}

The dataset collection process consists of two stages -- initial annotations are generated automatically using our proposed audio-visual method, and the annotations are then checked and refined by human annotators.

\subsection{Automatic pipeline}

The automatic computer vision pipeline to curate~\texttt{VoxConverse}
is similar to that used to compile \texttt{VoxCeleb1}~\cite{Nagrani17}
and \texttt{VoxCeleb2}~\cite{Chung18a}. 

% The pipeline for face detection and tracking is the same as in~\cite{Chung18a,Afouras19}. The tracked faces are then clustered into identities and 

\newpara{Stage 1.\  Collection of videos.}
The first stage is to obtain a list of videos. We start from a number of keywords including `panel debate' and `discussion' in order to obtain videos where multiple people are talking alternately or at the same time. 
The list of videos is obtained by searching the keywords on YouTube, and duplicate URLs that appear in the search results of multiple keywords is removed. Moreover, we remove the videos that are identical or very similar in content based on tf-idf features~\cite{teller2000speech} extracted from the YouTube auto-generated subtitles.
The list contains a range of videos,
 ranging from US presidential debates and talk shows to documentaries. 

\newpara{Stage 2.\  Shot detection.}
Shot boundaries are then determined to find within-shot frames for which face
tracking is to be run. The boundaries are found by comparing intensity and brightness
across consecutive frames~\cite{pyscenedetect}. 

\newpara{Stage 3.\  Face detection and tracking.}
A CNN face detector based on the
Single Shot Scale-invariant Face Detector (S3FD)~\cite{zhang2017s3fd}
is used to detect faces
 on every frame of the video. 
This detector allows the detection of faces at various scales and poses. 
Within each shot, face detections are grouped
together into face tracks using a position-based tracker, as in~\cite{Nagrani17,Chung18a}.
 
\newpara{Stage 4.\  Face-track clustering. }
A face recognition CNN is used to extract embeddings for every face track.
The network used here is based on the ResNet-50~\cite{He16} trained on the VGGFace2
dataset. The embeddings are extracted 5 times per face track at uniform intervals, and then averaged.
The embeddings are clustered using Agglomerative Hierarchical Clustering~\cite{day1984efficient}, but a large penalty is added to the distance matrix between overlapping face track so that they are never clustered together.

\newpara{Stage 5.\  Active speaker detection (ASD).}
The goal of this stage is to determine if the visible face is the speaker. 
Two systems are used for this purpose.
The first method uses a variant~\cite{chung2019perfect} of SyncNet~\cite{Chung17a}, which is a two-stream CNN that determines the active speaker by estimating the correlation between the audio track and the mouth motion of the video. 
The second method isolates the speech of the target speaker from a mixture of sounds using an audio-visual speech enhancement (AVSE) network~\cite{Afouras18} then uses an off-the-shelf voice activity detector, WebRTC~\cite{johnston2012webrtc}, to determine the speech segment.

Each method has its weaknesses -- the SyncNet ASD activates when the phoneme in the background speech matches the viseme shown on the target face since the model does not consider temporal context; the WebRTC voice activity detector is often activated from the residual signal left in the AVSE output which, despite the reduced power, causes false alarms.
Therefore, a face track is considered to be speaking only if both of the methods agree, which helps reduce false alarms from laughter and music.

\newpara{Stage 6.\  Labelling off-screen speech. } 
A pre-trained speaker recognition model~\cite{chung2020defence} is used to verify the identity of speech that comes from off-screen speakers. 
Any parts of the audio with positive voice activity detector (VAD) output, but with no visible active speaker is considered to be an off-screen speech segment.
Speaker embeddings are extracted for the whole video, then the off-screen speech segments are compared to all speech segments with visible active speaker using the cosine distance between the embeddings. If the cosine distance is below a threshold, the off-screen segment is assigned to the speaker; if not, the segment is left as unknown for the human annotator to verify. This procedure is closely related to the multi-modal diarisation method of~\cite{chung2019said}.
% \footnotetext{http://www.robots.ox.ac.uk/~vgg/data/voxceleb/competition2020.html}

\newpara{Discussion. }  
In creating the VoxCeleb datasets, very conservative thresholds were used in both active speaker detection and face verification, since it was necessary to be very certain about the speaker labels without any human intervention. This is, however, at the cost of high false rejection, which meant that a large number of true speech segments were discarded. 

In contrast, a speaker diarisation dataset must contain continuous audio recording with different identities speaking in turns. Therefore, we cannot discard parts of video based on low confidence, but the entire video must be labelled in full. The thresholds are optimised to minimise the overall Diarisation Error Rate (DER) (see section~\ref{sec:experiments}), since high false alarms and high false rejections both lead to increased man-hours during the manual verification and correction stage. Note how our pipeline consists of two ASD methods, as we show later, this redundancy is beneficial for performance.

\subsection{Manual verification}
 
The output of the automatic pipeline has been checked and corrected by the authors of this paper using a customised version of the VGG Image Annotator~\cite{dutta2019vgg,dutta2016via}. This was done so that the authors can identify the failure modes and make guidelines for the external annotators when the process is scaled up. The tool allows the user to watch and verify the annotation at various speeds, and with aid of video.
% See \url{http://www.robots.ox.ac.uk/~vgg/data/voxceleb2/spottheconv.mp4} for a video demonstration of the annotation process. 

During the annotation process, a number of failure modes were identified. The most common is non-visible speech segment assigned to the wrong speaker, but false alarm of the VAD and missed overlapped speech are also relatively common.

\newpara{Guidelines.}
Speech segments are split when pauses are greater than 0.25 seconds. Unlike some previous datasets in diarisation, laughter is not assigned to identities, as it is difficult to assign an accurate label to audience laughter. Anything that can be transcribed, including short utterances such as `yes' and `right', are considered to be speech. 
Known speakers are named in the annotation process to facilitate easier cross-checking. The annotators are asked to be as careful as possible that the marked boundaries are within 0.1 seconds of the true boundary.

\newpara{Quality check.}
In order to verify the quality of manually checked annotations, a subset of the data has been labelled independently by two different annotators. This subset contains 1 hour of material from 15 YouTube videos. The diarisation error rate between the two annotations is approximately 1\%, using the labels from one annotator as the reference and the other as the prediction. This error can be mostly attributed to disagreements on the source of off-screen speech segments. %The DER of the pipeline against the first annotation on this subset is 8\%. 

\newpara{Discussion.}
Diarisation labels for `in the wild' conversations are difficult to obtain. It is almost impossible to manually annotate the segments in our dataset without the video. Even with the video, it can take 10 times the video duration to annotate segments to satisfactory quality if \textit{starting from scratch}, particularly for many speakers. In contrast, the verification of our audio-visual method output takes around twice the video duration, and is possible with less experienced annotators. 

The time taken to annotate correlates strongly to the quality of the output from the automatic method. The first few videos in the development set were annotated with initial hyperparameters that gave relatively poor performance. The diarisation labels were then manually fixed, and the parameters were re-tuned on this data to minimise the diarisation error rate. More videos were then generated using the new set of parameters and this process was repeated a few times. While it is possible that some types of errors are more time-consuming for humans to fix compared to others, we have observed that the annotation became faster after each iteration.

%%% ========== ========== ==========
%%% Experiments
%%% ========== ========== ==========
\section{Experiments}
\label{sec:experiments}

We compare our audio-visual method to an audio-only DIHARD 2019 baseline, and also compare performance to two ablations.

\newpara{DIHARD 2019 baseline.}
The second DIHARD \cite{sell2018diarization, ryant2019second} challenge provides a baseline system based on the JHU submission of the first DIHARD challenge. We use this public code\footnote{https://github.com/iiscleap/DIHARD\_2019\_baseline\_alltracks} as an audio-only baseline. 

The overall procedure is as follows. Speech segments are obtained using VAD, and divided into short overlapping segments (1.5s with 0.75s overlap). Speaker embeddings are extracted using the x-vector~\cite{snyder2018x} system, and the similarities between the embeddings are scored with a pre-trained probabilistic linear discriminant analysis (PLDA) \cite{ioffe2006probabilistic, kenny2013plda} model also provided in the code.
Segments are then grouped using agglomerative hierarchical clustering (AHC) based on PLDA scores. We report the best performance by tuning the threshold of the AHC on the development set.

Two variants are compared, with and without the speech enhancement module~\cite{sun2018speaker} which has been made publicly
available\footnote{https://github.com/staplesinLA/denoising\_DIHARD18}. 
The system uses a Long short-term memory (LSTM)
based speech denoising model trained on simulated training
data. 
This model shows state-of-the-art performance on speech enhancement, and has shown its effectiveness for diarisation in the first DIHARD challenge.
% The enhancement process reduces the level of background noise such as music or babble noise, significantly reducing the VAD error. 

\newpara{Ablations.}
A crucial design choice that we made is that we used two active speaker detection methods, and a segment was only marked positive when both methods gave a positive output. We consider two ablations of our method -- one using only SyncNet-based ASD, and the other using only AVSE-based ASD.

\newpara{Evaluation protocol.}
Methods are evaluated on the \texttt{VoxConverse} development set.
We use the diarisation error rate (DER), defined as the sum of missed speech (MS), false alarm speech (FA), and speaker misclassification error (speaker confusion, SC). A forgiveness collar of 0.25 seconds is applied in order to compensate for small inconsistencies in annotation.

 \newpara{Training.} All thresholds are tuned on the \texttt{VoxConverse} development set.
The AHC threshold for speaker clustering is the only hyperparameter to be tuned in the audio-only baseline. 
The audio-visual method requires three key thresholds -- cosine distance for face clustering, SyncNet confidence for active speaker detection, and cosine distance for speaker identification. 
% ** isn't there also a threshold for VAD? ** WebRTC VAD has 3 settings, but even the most aggresive one is not aggressive enough.
The first of these affect performance the most, since any error in the identity clustering directly causes speaker confusion.
% \arshasays{Any numbers/graph to show this?} - no time or space

\begin{table}[t]
\centering
\renewcommand\arraystretch{1.2}
\setlength{\tabcolsep}{4pt}
\caption{
Results on the dev set using baseline methods and our proposed audio-visual method.
All values are in \%.
{\bf MS:} missed speech;
{\bf FA:} false alarm;
{\bf SC:} speaker confusion;
{\bf DER:} diarsation error rate (where $DER = MS + FA + SC$).
 For each metric, the lower the better. $\dagger$ Audio-only baselines.
}
\label{table:results}
\footnotesize
\begin{tabular}{ l |  r  r  r r  }
\toprule
%   & \multicolumn{4}{c|}{Development set}  & \multicolumn{4}{c}{Test set} \\
% \midrule
 \textbf{Name}   & \textbf{MS} & \textbf{FA}  & \textbf{SC} & \textbf{DER} \\ % & \textbf{MS} & \textbf{FA}  & \textbf{SC} & \textbf{DER} \\ 
\midrule
DIHARD 2019 baseline~\cite{sell2018diarization} $\dagger$                       & 11.1 & 1.4 & 11.3 & 23.8  \\ %     &  10.9 & 1.4 & 12.2 & 24.5          \\ 
DIHARD 2019 baseline w/ SE~\cite{sell2018diarization,sun2018speaker} $\dagger$  & 9.3 & 1.3 & 9.7 & 20.2   \\ %   &  9.3 & 1.3 & 9.6 & 20.1           \\ 
\midrule 
 Ours (SyncNet ASD only)        & 2.2 & 4.1 & 4.0 & 10.4    \\ % 	& 2.3 & 4.1 & 4.4 & 10.8   \\
 Ours (AVSE ASD only)           & 2.0 & 5.9 & 4.6 & 12.4	\\ %	& 1.9 & 5.4 & 4.9 & 12.2   \\
 \textbf{Ours (proposed)}	   & 2.4 & 2.3 & 3.0 & 7.7	    \\ %    & 2.3 & 1.8 & 3.5 & 7.6        \\ 
 \bottomrule
\end{tabular} 
\normalsize
\vspace{-15pt}
\end{table}
\newpara{Results.}
Table~\ref{table:results} shows the results of all the evaluations. Our audio-visual method obtains a DER much lower than the audio-only state-of-the-art baselines, showing the efficacy of using visual information for diarisation on this dataset. The ablation analysis for the ASD methods proves the effectiveness of using two active speaker detectors -- the combined method has a significant decrease in false alarm rate for only a small increase in missed speech. 

With regards to the difficulty of \texttt{VoxConverse}, we note that the DIHARD 2019 baseline obtains a DER of about 20\% on our dataset (Table~\ref{table:results}), and hence there is a lot of room for improvement. While this is lower than the 26\% that the same model achieves on the extremely challenging DIHARD development set (with ground truth VAD), we hypothesize that this difference may be attributed to the use of a 0.25-second forgiveness collar in our evaluation protocol.

%%% ========== ========== ==========
%%% Conclusion
%%% ========== ========== ==========
\section{Conclusion}
We have developed a high performance audio-visual algorithm for automated diarisation, and used it to generate a new speaker diarisation dataset, \texttt{VoxConverse}, from `in the wild' videos. The pipeline is fully scalable and effective across a range of domains.
\texttt{VoxConverse} currently contains 70 hours of annotated video, but we are in the process of scaling up. The data will be used in the second VoxCeleb Speaker Recognition Challenge in October 2020 and, after that, will be released publicly to the research community free of charge.

% \vspace{5pt}
\section{Acknowledgements.}
This work is funded by the EPSRC Programme
Grant Seebibyte EP/M013774/1. Arsha is funded by a Google PhD Fellowship.
Triantafyllos is funded by the UK EPSRC
CDT in Autonomous Intelligent Machines and Systems and
the Oxford-Google DeepMind Graduate Scholarship. We are also very grateful to Mihir Bhushan for assisting with annotation. 
\clearpage
\raggedbottom
\bibliographystyle{IEEEtran}
\bibliography{shortstrings,vgg_local,mybib,vgg_other}
\end{document}